\documentstyle[epsf,epsfig,rotate]{mn}
\newcommand{\be}{\begin{equation}}
\newcommand{\ee}{\end{equation}}

\newcommand{\apj}{{\it ApJ, }}

\newcommand{\icar}{{\it Icarus, }}

\title
[ MHD turbulence and a planet]
{The interaction of a giant planet with a disc with MHD turbulence II:
The interaction of the planet with the disc}
\author[R.P.Nelson \& J.C.B.Papaloizou]{Richard P. Nelson \&
John C.B. Papaloizou \\
Astronomy Unit, Queen Mary, University of London, Mile End Rd, London E1 4NS}

\date{Received/Accepted}

\begin{document}

\maketitle

\begin{abstract} 
We present a global
MHD simulation of a turbulent accretion disc interacting with a giant
protoplanet of $5$ Jupiter masses in a fixed circular orbit.
The disc model
had aspect ratio  $H/r=0.1$, and in the absence of the protoplanet
a typical value of the Shakura \& Sunyaev (1973)
stress parameter $\alpha =  5 \times 10^{-3}$.
As expected from previous work the protoplanet was found 
to open and maintain a significant gap in the disc,
with the interaction leading to inward migration of the protoplanet orbit
on the expected time scale.
No evidence for a persistent net mass flow through
the gap was found. However, that may be because 
an extensive  inner cavity could not be formed  for the model adopted.

Spiral waves
were launched by the protoplanet
and although these appeared to be diffused
and dissipated through interaction with the turbulence, they produced
an outward angular momentum flow which compensated for
a reduced flux associated with the MHD turbulence near the planet,
so maintaining the gap. 

When compared with laminar disc models,
with the same estimated $\alpha$ in the absence of the planet,
the gap was found to be deeper and wider indicating that the turbulent
disc behaved as if it in fact possessed a smaller $\alpha$, even though
analysis of the turbulent stress indicated that it was not significantly 
affected by the planet in the region away from the gap.
This may arise for two reasons. First,
unlike a Navier--Stokes viscosity with anomalous viscosity coefficient,
the turbulence does not provide
a source of constantly acting friction in the near vicinity of the planet that
leads to steady mass flow into the gap region.
Instead the turbulence is characterised by large fluctuations 
in the radial velocity, and time averaging of these fluctuations over 
significant time scales is required to recover the underlying mass flow
through the disc. In the vicinity of the planet the disc material
experiences high amplitude periodic perturbations on time scales 
that are short relative
to the time scale required for averaging. Consequently
the disc response is likely to
be significantly altered relative to that expected from a Navier--Stokes model. 
Second, the simulation indicates that
an ordered magnetic connection
between the inner and outer disc can occur
enabling  angular momentum to flow out across  
the gap, helping
to maintain it independently of the protoplanet's tide.
This type of effect may assist gap formation for smaller
mass protoplanets which otherwise would not be able to maintain
them.

There is also some evidence that magnetic connection between
the circumstellar disc and material that flows into the protoplanet's
Hill sphere may lead to significant magnetic breaking of the resulting
circumplanetary disc, thereby modifying the expected gas accretion rate onto
forming gas giant planets.

\end{abstract}


\section{Introduction}\label{S0} 
\noindent

The recent and ongoing discovery of extrasolar giant planets has stimulated
renewed interest in the theory of planet formation 
(e.g. Mayor \& Queloz 1995;
Marcy, Cochran, \& Mayor 1999; Vogt et al. 2002). At the present time
there are two main competing theories for the formation of 
giant gas--rich planets,
both of which assume formation in a protostellar disc.
In the first, giant planets are envisaged 
to form directly from the gravitational
fragmentation of a young protostellar disc (e.g. Boss 2001). In the second,
giant planets form by the gradual build--up of a rocky core through coagulation
of solids, and this core undergoes significant gas accretion once the core mass
reaches $\simeq 15$ Earth masses (e.g. Bodenheimer \& Pollack 1986; 
Pollack et al. 1996). In either scenario, the gravitational
interaction between the protoplanet and the protostellar disc from which it 
forms will play an important role in subsequent evolution.

There have been a number of studies of disc--protoplanet interaction.
In the standard picture a protoplanet exerts torques on a protostellar disc
through the excitation of spiral density waves at Lindblad resonances
(e.g. Goldreich \& Tremaine 1979),
and these waves carry with them an associated angular momentum flux.
This angular momentum is deposited in the disc material
where the 
waves are damped, leading to an exchange of angular momentum
between protoplanet and disc. The planet exerts a positive torque
on the disc that lies exterior to its orbit and a negative torque on
the interior disc. 

Previous studies of the interaction between a protoplanet and a viscous
disc (Papaloizou \& Lin 1984; Lin \& Papaloizou 1993) indicate that a
sufficiently massive protoplanet can open up an annular gap in the disc 
about the planet's orbital radius provided: {\it i}) the tidal torque due to
the planet is greater than the internal viscous torque; {\it ii}) the spiral
density waves excited by the planet form shocks, ensuring that the torque
exerted by the planet is applied locally to the disc (Ward 1997). 
For most protostellar disc models the protoplanet 
needs to be approximately a Jovian
mass for gap formation to occur.
Recent simulations (Bryden et al. 1999; Kley 1999; Lubow, Seibert, \& 
Artymowicz 1999; 
D'Angelo, Henning, \& Kley 2002)
examined the formation of gaps by giant protoplanets,
and also estimated the gas accretion rate onto the planets. The detailed
form of
the gaps, and by association the gas accretion rate onto the protoplanet, is 
found to be a function of the viscosity in the disc, as well as the planet
mass and the disc vertical scale height, $H$. The orbital evolution
of a Jovian mass protoplanet embedded in a standard viscous protostellar
disc model 
was studied by Nelson et al. (2000), who showed that gap formation 
leads to the creation of a low density inner cavity in which the planet 
orbits, and that continued interaction with the outer disc leads to inwards
migration on a time scale of a few $\times 10^5$ yr. It was estimated that
a Jovian mass protoplanet would increase its mass to $\sim 3$ Jupiter masses
by accretion of gas from the disc for this time. The disc models in these
studies all modelled the disc viscosity through the Navier--Stokes equation,
using an $\alpha$ model for the kinematic viscosity.

Until quite recently most models of viscous accretion discs used the
Shakura \& Sunyaev (1973) $\alpha$ model for the anomalous
disc viscosity. This assumes
that the viscous stress is proportional to the thermal pressure in the disc
without specifying the origin of the viscous stress (but assumed to arise from
some form of turbulence). Work by Balbus \& Hawley (1991) indicated that
significant angular momentum transport in weakly magnetised discs could arise 
from the magnetorotational instability (MRI -- or the Balbus--Hawley 
instability).
Subsequent non linear numerical simulations performed using a local
shearing box formalism
(e.g. Hawley \& Balbus 1991; 
Hawley, Gammie, \& Balbus 1996; Brandenburg et al. 1996)  confirmed
this and showed that the saturated non linear outcome of the MRI
is MHD turbulence
with an associated viscous stress parameter $\alpha$ of between $\sim 5 \times
10^{-3}$ and $\sim 0.1$, depending on the initial magnetic field configuration.
More recent global simulations of MHD turbulent discs
[e.g. Armitage (1998); Hawley (2000); 
Hawley (2001); Steinacker \& Papaloizou (2002)] confirm the picture 
provided by the local shearing box simulations. 

In a companion paper to this one (Papaloizou \& Nelson (2002) -- hereafter
paper I) we present the results of 3D global MHD simulations of 
turbulent discs initiated with zero net flux magnetic fields. We examine
the behaviour of the discs as a function of varying the disc scale height, $H$,
the extent of the radial domain, the numerical resolution, and the initial
magnetic field configuration (poloidal or toroidal). We find that these disc
models produce volume averaged values for the stress parameter 
$\alpha \sim 5 \times 10^{-3}$. We note that this value is typical
of that expected in protostellar discs to maintain
the canonical mass accretion rate of $10^{-8}$ M$_{\odot}$ yr$^{-1}$ onto
T Tauri stars (e.g. Hartmann et al. 1998). We also consider the
issue of time averaged estimates for the radial distribution of
the turbulent quantities such as the time averaged stress parameter
$\alpha$ and the time averaged mass flux through the disc. Balbus \&
Paploizou (1999) noted the need to consider time averages of these
quantities in order to connect the behaviour of MHD turbulent discs models
with standard viscous accretion disc theory. Indeed we find   reasonable
agreement
between the expectations of viscous disc theory and the results
of the simulations in paper I, when appropriate time averages
are considered. One of the models presented in paper I (model E) was used as
the basis for the work presented in this paper.

The calculations presented in this paper and in paper I assume that
the equations of ideal MHD apply. Questions have been raised concerning the
degree of ionisation in protostellar discs (e.g. Gammie 1996), and thus about
the applicability of the equations of ideal MHD in the modelling of
protoplanetary discs. This is a complex issue that has yet to be fully 
resolved, so we assume that the equations of ideal MHD are a valid approximation
in these first papers on the interaction between turbulent discs and
embedded protoplanets. Future work will address the issue of the ionisation
fraction in protostellar discs, and the effects this may have on disc--planet
interactions.

Significant work has already been done on examining the interaction between
giant protoplanets and viscous protostellar discs using laminar
 $\alpha$ disc models.
The improvement and availability of high performance computing
resources has now made it possible to develop global models of
MHD turbulent discs in which the underlying mechanism responsible
for angular momentum transport is explicitly calculated. 
The next important step in understanding the interaction between 
protostellar accretion discs and embedded giant protoplanets
is to compute models of turbulent MHD discs interacting with giant
protoplanets. This is the subject of the present paper.

The underlying disc model that we consider (described in detail in paper I)
has a constant  ratio of disc scale height to radius of $H/r=0.1$, and
a volume  and time
averaged value of the stress parameter $\alpha =  5 \times 10^{-3}$.
In order for a protoplanet to form a gap in such a disc, it must
have a mass of at least 3 Jupiter masses (assuming the central stellar mass is 
1 M$_{\odot}$) according to the thermal criterion (see section~\ref{Sgap}).
In order to be certain of generating a clean gap we consider a protoplanet
with 5 Jupiter masses. We note that at the present time it is beyond our
computational resources to consider thinner discs and lower mass planets.
Protostellar discs are more typically expected to have $H/r=0.05$, but
to model such discs would require a doubling of the number of grid cells
in each direction. We have found from conducting numerical experiments
that zero net magnetic flux disc models require a numerical resolution
similar to that we
consider here in order to sustain MHD turbulence at a reasonable level, and to 
generate reasonable values of $\alpha$. Ideally we would wish to perform
simulations of discs with $H/r=0.05$ and Jovian mass protoplanets.
However, we believe that the results obtained
will still provide a reliable picture of how protoplanets
interact with turbulent discs.

In particular,
as expected from previous work, the protoplanet was found
to open and maintain a significant gap in the disc,
with the interaction leading to inward migration of the protoplanet orbit
on the expected time scale.
Spiral waves
were launched by the protoplanet
producing
an outward angular momentum flow and reducing that
provided by the MHD turbulence near the planet.
The turbulent
disc behaved as if it possessed a smaller $\alpha$
than expected.
The characteristic flow induced by the protoplanet
was found to
facilitate
an ordered  magnetic connection
between the inner and outer disc
enabling  angular momentum to flow  out across
the gap. This effect provides an additional
mechanism for maintainin the gap and may account
for the apparently smaller $\alpha$, along with the fact that the turbulence 
does not provide a source of constantly acting friction, unlike a
Navier--Stokes viscosity.

The plan of the paper is as follows. In sections~\ref{S1} and \ref{S2} 
we described our initial
model set up and numerical procedure. In section~\ref{S3} we discuss the issue of time averaging the relevant disc quantities and the connection to
standard viscous disc theory. In section~\ref{Sgap} we discuss the basic ideas
of disc--protoplanet interactions, and our expected results based
on this previous work. In section~\ref{evolution} we discuss the time dependent
evolution of the model, and in section~\ref{compare} we compare
our results with those obtained from laminar viscous disc models.
Finally in section~\ref{conclusions} we discuss our results and draw our conclusions.

\section{Initial model setup} \label{S1} 
\noindent
The governing equations for MHD written in a frame rotating
with a  uniform  angular velocity $\Omega_p {\bf {\hat k}} $
with $ {\bf {\hat k}} $ being the unit vector in the vertical direction
are:
\begin{equation}
\frac{\partial \rho}{\partial t}+ \nabla \cdot {\rho\bf v}=0, \label{cont}
\end{equation}
\begin{eqnarray}
\rho \left(\frac{\partial {\bf v}}{\partial t}
 + {\bf v}\cdot\nabla{\bf v}\right) + 2\Omega_p {\bf {\hat k}}{\bf \times}{\bf v} & = &
-\nabla p -\rho \nabla\Phi \nonumber \\ & & +
\frac{1}{4\pi}(\nabla \times {\bf B}) \times {\bf B}, \label{mot}
\end{eqnarray}
\begin{equation}
\frac{\partial {\bf B}}{\partial t}=\nabla \times ({\bf v} \times {\bf B}).
\label{induct}
\end{equation}
where ${\bf v}, P, \rho, {\bf B}$ and $\Phi$ denote the fluid
velocity, pressure, density, magnetic field, and potential, respectively.

The angular velocity $\Omega_p$ corresponds to
that of a planet in fixed circular orbit at radius $D.$
The perturbing planet thus appears stationary in the chosen reference frame.
The potential $\Phi$ contains contributions due to gravity
and the centrifugal potential $-(1/2)\Omega_p^2 r^2.$
We use a  locally isothermal equation of state
\begin{equation}
P(r)= c(r)^2 \cdot \rho,
\end{equation}
where $c(r)$ denotes the sound speed
which is specified as a fixed function of $r.$

The disc model investigated may be described
as a cylindrical disc (e.g. Hawley 2001) and was considered
prior to the introduction of a perturbing planet in paper I as model E.
This disc model was such that $c(r)/r \Omega(r)=0.1$, so that the effective
ratio of disc scale height to radius $H/r=0.1$, and had a volume averaged value
of the stress parameter $\alpha \simeq 5 \times 10^{-3}$.
We adopt cylindrical coordinates $(z,r,\phi)$ 
based on the  centre of mass of the central star plus protoplanet
system. The number of grid cells used was $(N_r, N_{\phi}, N_z)=
(370, 580, 60)$. The gravitational
potential is due to the central star and 
internally orbiting planet. 
Thus we have
\begin{eqnarray}
\Phi & = &-\frac{GM_*}{|{\bf r} - {\bf r}_*| } - 
\frac{G m_p}{\sqrt{r^2+D^2 -2rD\cos(\phi-\phi_p) +b^2}} \nonumber \\
& &   - \frac{1}{2}\Omega_p^2 r^2.
\end{eqnarray}
The first term represents
the gravitational potential due to the central star,  
with $G$  being the gravitational constant and $M_*$ being the stellar mass. 
The second term is the gravitational potential due to a planet
in circular orbit at radius $D$ and azimuthal angle $\phi_p$ with a mass $m_p$.
The final term represents the centrifugal potential.
To avoid a singularity in the planet potential
we adopt a softening parameter $b=R_H/2$ where $R_H=D(m_p/3M_*)^{1/3}$ is
the Hill sphere radius of the protoplanet. This enables a
Roche lobe filling atmosphere to be set up
around the protoplanet before the gap construction
process is completed. Material flowing in  the neighbourhood of the protoplanet
then has to  move around it and  cannot be accreted.  We  thus cannot
simulate the accretion
of gas onto the protoplanet in this work. In previous work on
disc--planet interactions (e.g. Nelson et al. 2000), accretion
of gas by a protoplanet is modelled my simply removing matter from
within the Roche lobe at a specified rate. A technical difficulty arises
in a simulation like the one presented here that includes magnetic
fields, since one also has to consider what happens to the magnetic
field when gas is accreted by the protoplanet. This is the primary reason
why we do not explicitly calculate the gas accretion rate onto the protoplanet
in this work.

In this paper we consider the time dependent evolution
of a  turbulent disc together with a 
planet with $m_p$ equal to $5$ Jupiter masses.
We initially immerse the planet at $r \equiv D=2.2$ and $\phi_p=\pi$
into a fully turbulent
disc model covering the full azimuthal domain of $2\pi$.
The initial model we adopt is model E that
has been described in paper I.
When a planet is immersed directly into such a model
we found that large scale transients occurred which prevented the
efficient and quick formation of a gap. It was not possible to follow
such transients to a conclusion in a run of reasonable duration.
To avoid such problems we carved out a gap in the density profile centred
on the planet's radial location
before inserting the planet. This had the effect
of perturbing the  disc 
from a steady state, so it was allowed to relax
until a steady state had been reattained. The planet was then inserted
into the disc and the calculation initiated with the time $t$ being reset 
to zero.
We comment that in carving out the initial  gap the magnetic field
was not altered.
This means that the magnetic energy in the initial gap
would be significantly larger than expected for relaxed turbulence
under conditions of zero net flux.
However, the runs were subsequently continued for about $100$
planet orbits which is significantly longer than the time required
for adjustment of the local turbulence (see paper I).
Further, after such times, the magnetic stresses appeared
to have reached a steady pattern and to be largely associated
with magnetic field concentrated into wakes which were absent 
when the planet was initially inserted. Thus the behaviour
of the magnetic field in the gap region is not determined
by the conditions prevailing when the gap was set up. 

A test run was performed in the absence of an embedded protoplanet
to ensure that the turbulent disc relaxed to a state with the expected
magnetic energy once an initial gap had been carved out. 
The results of this run showed that the disc returns to its normal turbulent 
state within a few orbits.

The form of the azimuthally averaged density profile
located at the vertical midplane of the disc
at the stage when the planet is introduced
is plotted in figure~\ref{fig1}.

As in paper I, units
are adopted such that $r_{1} =1 $ and $GM=1$, where $r_1$ is the radius
of the inner radial boundary of the computational domain. 
We note that a single simulation of this type requires of order 30,000 CPU
hours on an Origin 3800 parallel processing facility.

\subsection{Numerical procedure} \label{S2}
\noindent
The numerical scheme that we employ is
based on a spatially second--order accurate method that computes the
advection using the monotonic transport algorithm (Van Leer 1977).
The MHD section of the code uses the
method of characteristics
constrained transport (MOCCT) as outlined in Hawley \& Stone (1995)
and implemented in the ZEUS code.
The code has been developed from a version
of NIRVANA originally written by U. Ziegler (Ziegler \& Rudiger 2000).

\section{Numerical results} \label{S3a}

Before describing results in detail we
discuss some average quantities used in their description.
 
\subsection{Vertically and horizontally averaged stresses,
 angular momentum transport and external torques}\label{S3}
\noindent
For a global  coarse grained description of the flow, as in paper I, we
use quantities that are both
vertically and azimuthally averaged over the $(\phi, z)$ 
domain (e.g. Hawley 2000)
and in some cases an additional time average. Adopting the same
formalism and notation as in paper I, we write the
vertically and azimuthally averaged
continuity equation  in the form
\begin{equation}
\frac{\partial \Sigma}{\partial t}+\frac{1}{r}
\frac{\partial\left(r\Sigma\overline{v_r}\right)}{\partial r}=0, \label{cona}
\end{equation}
where
\noindent the  disc surface density is given by
\begin{equation}
\Sigma = {1\over 2\pi}\int \rho dz d\phi.
\end{equation}

We adopt the Shakura  \& Sunyaev (1973)
$\alpha$ stress parameter appropriate to the
total stress as  defined  in paper I.
Then the  vertically and azimuthally averaged azimuthal
component of the equation of motion can be written in the form
\begin{equation}
\frac{\partial \left(\Sigma \overline{j}\right)}{\partial t}
+\frac{1}{r}\left(
\frac{\partial\left( r\Sigma\overline{v_r}\overline{j}\right)}{\partial r}
+\frac{\partial\left(\Sigma r^2\alpha\overline{P /\rho}\right)}{\partial r}
\right) =-\int \rho
 {\partial \Psi \over \partial \phi} {d\phi dz\over 2\pi} .\label{mota}
\end{equation}
Here $j=rv_{\phi}$ is the specific angular momentum and the
term on the right hand side represents the azimuthally and 
vertically integrated 
torque acting on the disc (absent in paper I)  which is here due to the internally orbiting planet.
Note that this form is identical
to that obtained in standard viscous disc theory (see Balbus \& Papaloizou 1999).

\noindent Equations (\ref{cona}) and (\ref{mota})  may  be used
with or without a time averaging procedure in addition to
horizontal and vertical averaging (see paper I).

\subsection{Disc--Planet Interactions and Gap Formation} \label{Sgap}
In the usual picture of disc--protoplanet interactions, the presence of a 
protoplanet orbiting in a gaseous accretion disc leads to the excitation
of spiral density waves at Lindblad resonances, which propagate from the 
locations where they are launched (e.g. Goldreich \& Tremaine 1979). These
spiral waves carry with them an associated angular momentum flux that is 
deposited in the disc where the waves are damped, either through shocks
or viscous effects.

Previous work on the interaction between a giant protoplanet
and a viscous protoplanetary disc indicates that an annular
gap is expected to form in the vicinity of the protoplanet provided
that two criteria are satisfied (e.g. Lin \& Papaloizou 1993).
The first criterion (the so--called viscous criterion) requires that
\begin{equation} 
\frac{m_p}{M_*} > 40 \alpha \left(\frac{H}{R} \right)^2 
\label{visc_gap}
\end{equation}
which arises from the requirement that tidal torques induced by the
protoplanet exceed viscous torques in the disc.
The second criterion (the so--called thermal criterion) requires that
\begin{equation}
\frac{m_p}{M_*} > 3 \left(\frac{H}{R}\right)^3
\label{therm_gap}
\end{equation}
and comes from the requirement that the spiral waves introduced into
the disc flow by the planet be non linear so that spiral shocks form
(Ward 1997).

The disc model that we employ in this work is characterised by
a value of $H/R=0.1$ and a volume averaged value of $\alpha \simeq 0.005$
(see paper I).
It is thus expected that the presence of a 5 Jupiter mass protoplanet
will lead to gap clearing. Previous numerical work on disc protoplanet
interactions (Nelson et al. 2000) suggests that gap formation
in a disc in which the gas can accrete onto a central star will 
lead to the eventual formation of a low density inner cavity. We note that
our use of an inner boundary layer in this work will prevent an extensive
 cavity from forming.

\subsection{Time dependent evolution of the model} \label{evolution}

\begin{figure}
\centerline{
\epsfig{file= 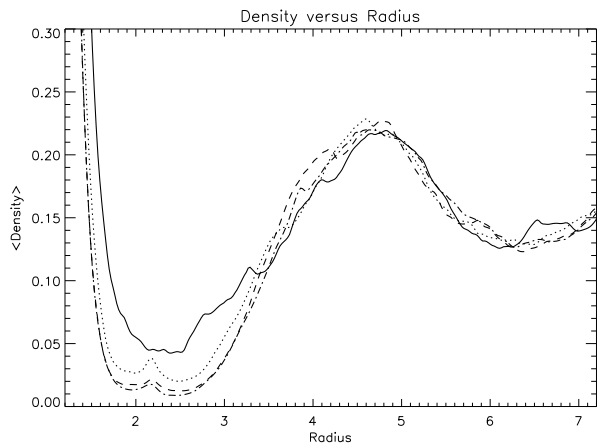,width=\columnwidth} }
\caption[]
{The azimuthally averaged density at the vertical midplane of the disc
is plotted as a function of radius 
at $t=0$ (solid line). The remaining curves correspond to the same
physical quantity after times of $t=675.6$ (dotted line), $t=1351.2$ 
(dashed line), and $t=2047.28$ (dot-dashed line). These times correspond
to 0, 33, 66, and 100 orbits of the protoplanet, respectively.
Note that the initial gap has been deepened and widened considerably.
Also some accretion into the central regions has occurred.
}
\label{fig1}
\end{figure}

The disc model together with planet was run for $2040$ time units.
This corresponds to $\sim 324$ orbits at $r=1$, about $100$ orbits at
$r=2.2$ (where the planet is located), and about 17 orbits at $r=7.2$.
We found that the initial gap considerably deepened and 
accretion onto the central parts
occurred producing something like a central cavity seen in
previous work (e.g. Nelson et al. 2000). We note that this is unable
to   become very 
extensive in the present work due to our adoption of an inner boundary layer
 which prevents  mass from flowing through the inner boundary. The
behaviour of the azimuthally averaged value of $\rho(r)$ at the disc midplane
throughout the run is indicated in figure~\ref{fig1}.

Throughout the run MHD turbulence is sustained in regions
of the disc not strongly affected by the planet.
However, the planet has a large enough mass to strongly perturb the disc locally
generating outward and inward propagating density waves as well as extending
the initial gap as expected from the discussion in section~\ref{Sgap}.

\begin{figure}
\centerline{
\epsfig{file= 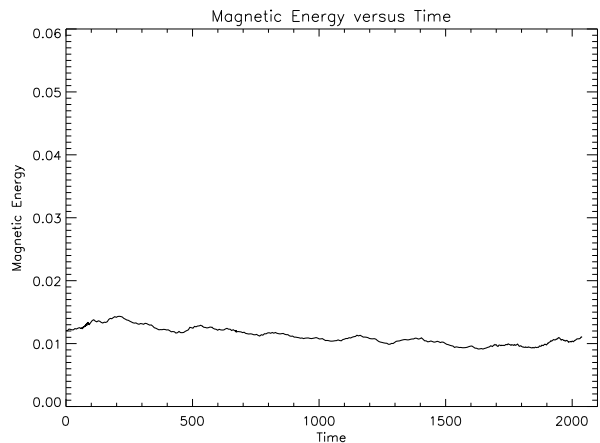,width=\columnwidth} }
\caption[]{Magnetic energy in the Keplerian domain
expressed in units of the volume integrated pressure
as a function of time. Note that the value of $\sim 0.01$ is
very similar to that obtained in the absence of the protoplanet.}
\label{fig2}
\end{figure}
\noindent

The magnetic energy in the Keplerian domain
expressed in units of the volume integrated pressure
is given 
as a function of time in figure \ref{fig2}.
A value of $\sim 0.01$ is maintained throughout.
This is similar to what is found for models without the planet, and
indicates that the perturbing presence of the protoplanet does not have
a strong effect on the dynamo operating {\em globally} within the disc.
However, there is evidence that the protoplanet affects the turbulence locally,
creating an ordered field where the material passes through the spiral shocks
in the gap region.

\begin{figure}
\centerline{
\epsfig{file=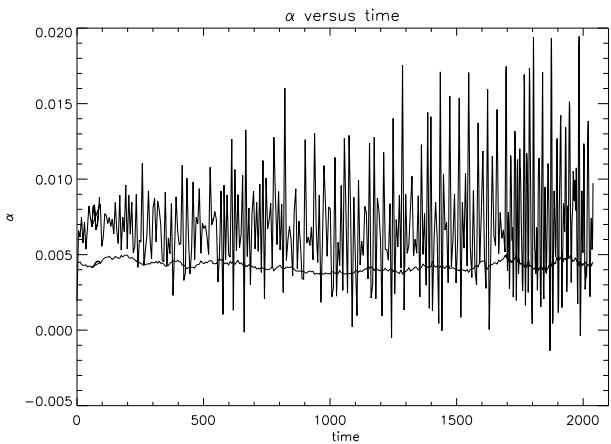,width=\columnwidth} }
\caption[]{The stress parameter
$\alpha$ volume averaged over the Keplerian domain is plotted
as a function of time.
The upper curve is  derived from 
the total stress while the lower curve   is derived from the
magnetic contribution.}
\label{fig3}
\end{figure}

The stress parameter
$\alpha$ volume averaged over the Keplerian domain is plotted
as a function of time  in figure~\ref{fig3}.
The magnetic contribution $\sim 0.004$ is similar to that occurring
without the planet. However, the typical total stress is much larger,
reaching peak values of $\alpha \sim 0.02$
This is due to the large contribution of Reynolds' 
stresses associated with the spiral waves launched by the protoplanet.

\begin{figure}
\centerline{
\epsfig{file=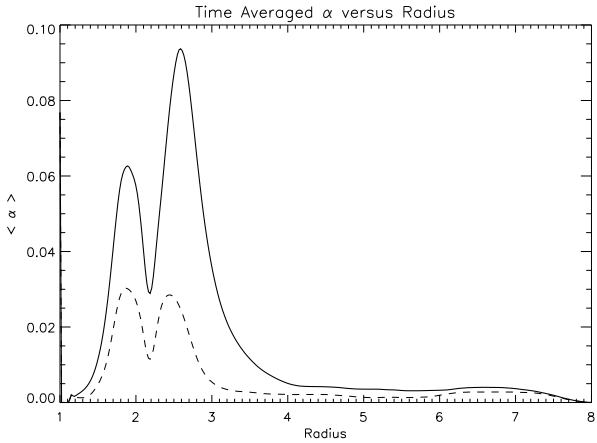,width=\columnwidth} }
\caption[]
{A time average of the vertically and azimuthally averaged
value of the stress parameter $\alpha$ plotted as a function of dimensionless
radius.
The solid curve corresponds to the combined effects of magnetic and
Reynolds' stresses whereas the dashed curve gives the contribution 
from magnetic stresses alone.
The average is taken between time $t=1650.0$
and $t= 2040.0$.
}
\label{fig4}
\end{figure}

\begin{figure}
\centerline{
\epsfig{file=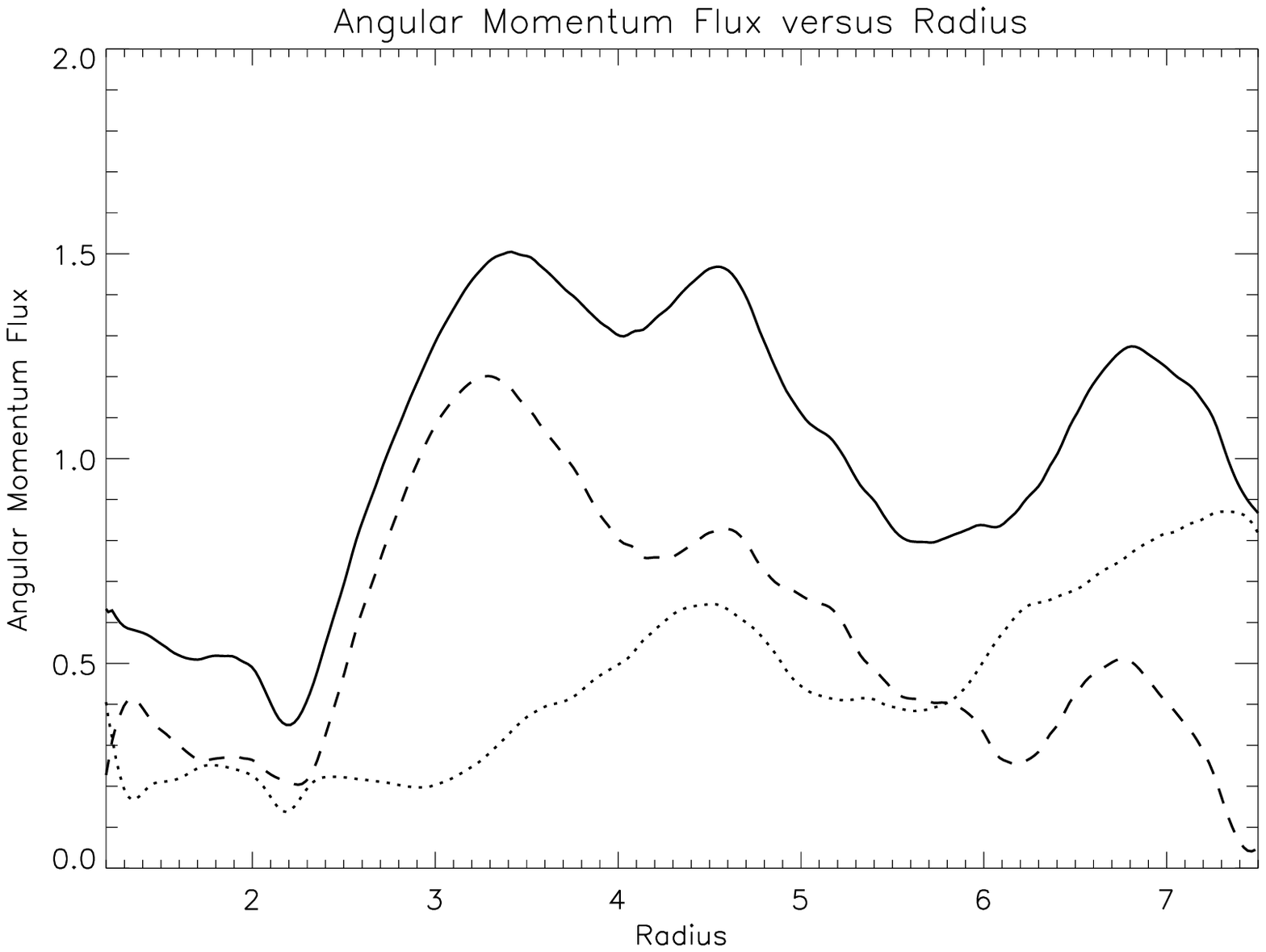,width=\columnwidth} }
\caption[]
{Time averages of the vertically and azimuthally averaged
values of the non advected
part of the angular momentum flux in arbitrary units
are plotted as a function of dimensionless radius.
The solid line gives the total contribution from the magnetic plus Reynolds'
stress.
The  
dashed line gives the  Reynolds stress contribution 
and the dotted line gives the magnetic stress contribution.
The  average is taken between  time $t=1650.0$
and $t=1920.0$.
}
\label{fig5}
\end{figure}

As in paper I we explore the 
behaviour of time averages of the
stress parameter $\alpha$ and its variation with radius.
Only in this way can a connection with earlier theories
of gap formation in viscous discs be made.
A time average
of the stress parameter
$\alpha$ 
taken between times $t=1650.0$
and $t= 2040.0$
is plotted as a function of dimensionless radius
in figure~\ref{fig4}.
Both the total stress parameter $\alpha$
and the magnetic contribution to it
become large in the vicinity of the planet $r < 3$ where the gap is.
This arises even though the actual magnetic stress $T_m = B_r B_{\phi}/ 4 \pi$
decreases in
the gap region once the gap is formed,
as indicated by figure~\ref{fig5}, because the pressure $P$
in the gap decreases by an even larger fraction (note that magnetic stress 
parameter $\alpha = -T_m/P$).
Further away beyond $r>$ 5 -- 6, values similar to those pertaining
to the disc without a planet are attained, indicating that the planet does
not have a significant effect on the turbulence outside of the gap region.

\begin{figure}
\centerline{
\epsfig{file=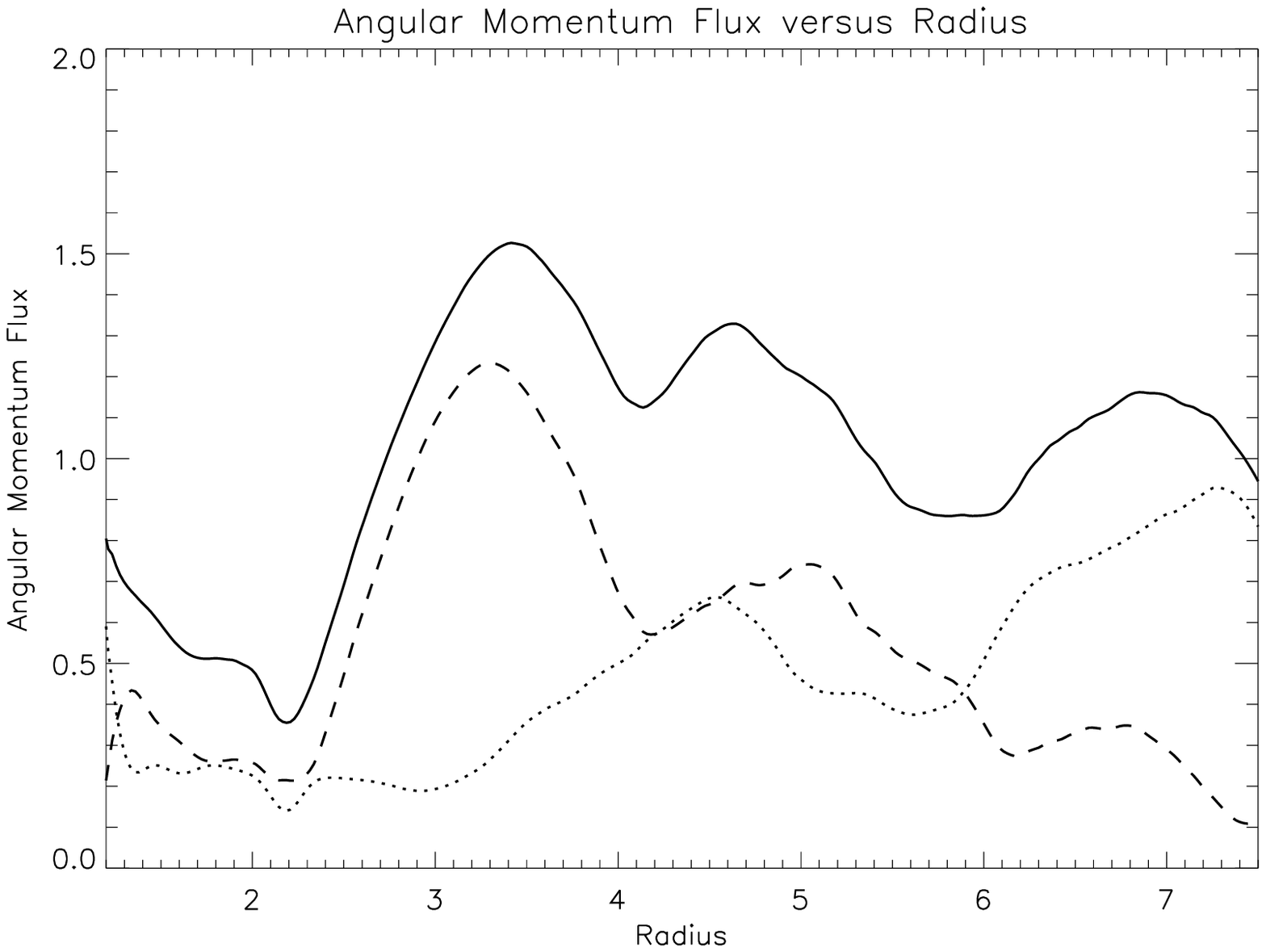,width=\columnwidth} }
\caption[]
{Time averages of the vertically and azimuthally averaged
values of the non advected part of the
angular momentum flux in arbitrary units
are plotted as a function of dimensionless radius.
The solid line gives the total contribution from the magnetic plus Reynolds'
 stress.
The 
dashed line gives the Reynolds stress contribution
and the dotted line gives the magnetic stress contribution.
The average is taken between  time $t=1650.0$
and $t=2040.0$.
}
\label{fig6}
\end{figure}

Inspection of equation
(\ref{mota}), however, suggests we should consider the 
non advected part of the angular momentum
flux which is proportional to $r^2\times{\rm stress}$
or 
$\Sigma r^2\alpha\overline{P /\rho}$ when discussing the disc evolution.
We plot time averages of the angular momentum fluxes in
figures \ref{fig5} and \ref{fig6}.
The time averages for these are over the time intervals
$1650 <t<1920$ and $ 1650 <t<2040$ respectively
and they both show the same behaviour indicating a
stable pattern of behaviour has been established
relatively quickly.

This behaviour may be understood in the context of gap
formation theory as applied to laminar discs
with Navier--Stokes viscosity (e.g. Bryden et al. 1999).
Consider equation (\ref{mota}) under the assumption
of a disc in a quasi steady state with ${\overline v_r} =0.$
Suppose also there is a completely empty gap in which the planet is located.
Then equation (\ref{mota}) predicts that the angular momentum flux
tends to a constant value at large distances from the planet.
The flux is zero within the gap where there is no material,
and is generated by the 
tidal torque term at the disc edge near the planet.

\begin{figure}
\epsfig{file=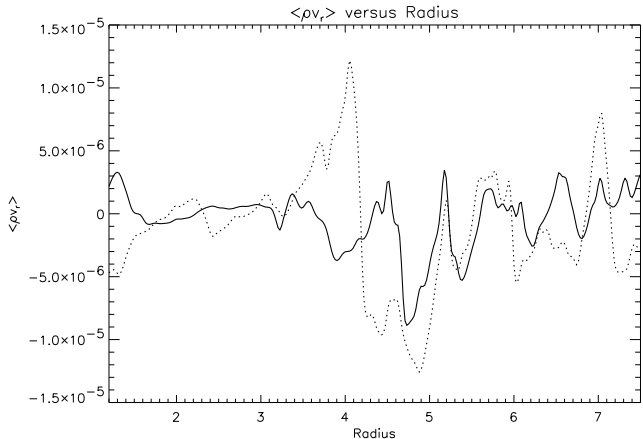,width=\columnwidth} 
\caption[]
{ Time averages of the product of the
vertically and azimuthally averaged
values of $\Sigma/L_z$ and $v_r$ in arbitrary units
are plotted as a function of dimensionless radius.
For the solid curve the  average is taken between time $t=1650.0$
and $t=2040.0$ while for the dotted curve
it is taken between $t=1650.0$
and $t=1920.0$ .
}
\label{fig7}
\end{figure}

The behaviour of the fluxes illustrated in figures \ref{fig5}
and \ref{fig6} can be related to this expectation.
We see that the total flux dips to a minimum at the location
of the planet orbit and rises to an asymptotic value at large 
distances. The contribution from the Reynolds' stresses initially
rises, representing the generation of spiral waves by the protoplanet in
the region of the gap wall and beyond,
and reaches a maximum
close to the outer 1:2 Lindblad resonance location, beyond which there is
no further generation of spiral waves by a planet on a circular orbit.
Beyond this region the contribution from the Reynolds' stresses decreases
as these waves are dissipated. The magnetic contribution rises from a small
but non zero value near the planet to make most of the contribution
at large radii. There the behaviour is that of an unperturbed turbulent disc.
The persistence of the magnetic stresses through the gap
may be significant in that this can allow for the transmission
of angular momentum from one side of the gap to the other
without needing the contribution of a tidal torque from the 
planet. This may enhance the gap formation process
with respect to what may be expected from non magnetic laminar disc theory
(see section~\ref{compare}).

The angular momentum
fluxes plotted in figures~\ref{fig5} and \ref{fig6}
indicate a net outward value supplied by the planet's tide.
This results in inward migration as has been verified by
direct calculation of the torques acting on the planet itself.
The torque, ${\dot J}$, arising from the disc beyond the planet orbital radius
indicates an inward migration time defined by
$J_{planet}/{\dot J} \simeq 2 \times 10^4$ planetary orbits,
for a disc mass
normalisation such that there exist two Jupiter masses interior to the
orbital radius of the planet $r_p=2.2$. Interestingly, the viscous timescale
at the planet location defined by the usual expression $t_{\nu} = r^2/(\alpha
H^2 \Omega) \simeq 1.03 \times 10^4$ planetary orbits. It is clear that
the planet undergoes inward migration at a rate that is in broad
agreement with the expectations of type II migration
theory [Lin \& Papaloizou (1986, 1993); Nelson et al. (2000)].

As in the unperturbed turbulent disc described in paper I, the radial velocity
shows large temporal fluctuations about a small
mean value. However, in this case when the planet is present
the mean values tend to be smaller giving some indication that
planetary tides balance viscous transport.
Time averages of the product of the
vertically and azimuthally averaged
values of $\Sigma/L_z$
and $v_r$, in arbitrary units,
are plotted as a function of dimensionless radius in figure~\ref{fig7}.
Here $L_z$ is the
extent of the vertical domain.
The averages  are taken between  time $t=1650.0$
and $t=1920.0$
and between $t=1650.0$
and $t=2040.0$.
Because of the small values the two averages show significant deviation.
Nonetheless the net mass flow is small.
This is particularly the case interior to the planet where there
is always a zero value at some radius
near to $r=D.$ That indicates no flow from exterior
the planet to the interior for the duration of
the averaging interval. Thus while there is material within the gap,
we see no evidence of a net flow through in this run.
In this context we note that results from laminar disc theory
produce models in which
significant mass flow through gaps may occur, as
has been recently  emphasized by Lubow, Seibert,
\& Artymowicz (1999). Our results suggest that different behaviour
may be exhibited in an MHD turbulent disc, and indicate that the rate of
mass accretion onto gap forming protoplanets may be be smaller
than is observed in laminar disc models.

\begin{figure*}
\centerline{
\epsfig{file=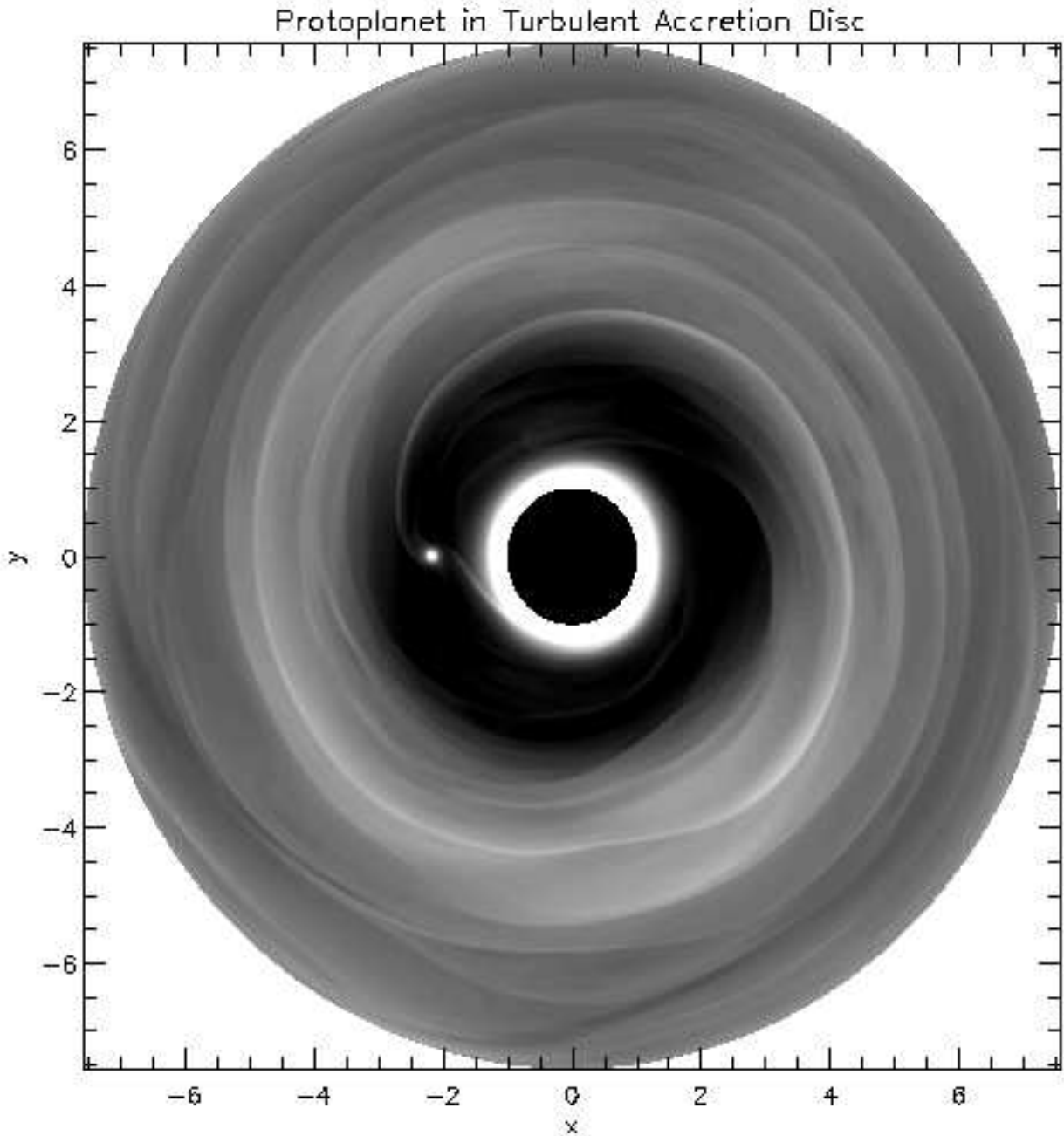,width=\textwidth} }
\caption[]
{ A typical contour plot for the midplane disc density at 
time $t=1680.0.$
The gap, planet and wakes it causes in the disc
are apparent. The underlying turbulence has the effect of slightly
blurring this picture relative to those obtained from simulations performed
using laminar accretion discs   with anomalous Navier--Stokes viscosities.
}
\label{fig8}
\end{figure*}

The density distribution in the gap region resembles
that obtained from laminar disc theory. A contour plot
of the midplane density
is given in figure \ref{fig8}. The wakes induced by the planet
through the excitation of spiral waves are clearly visible.
However in this case because of the turbulence,
they appear much less sharply defined than in the laminar
disc case (see Nelson et al. 2000 and figure~\ref{fig15b}).

\begin{figure*}
\centerline{
\epsfig{file=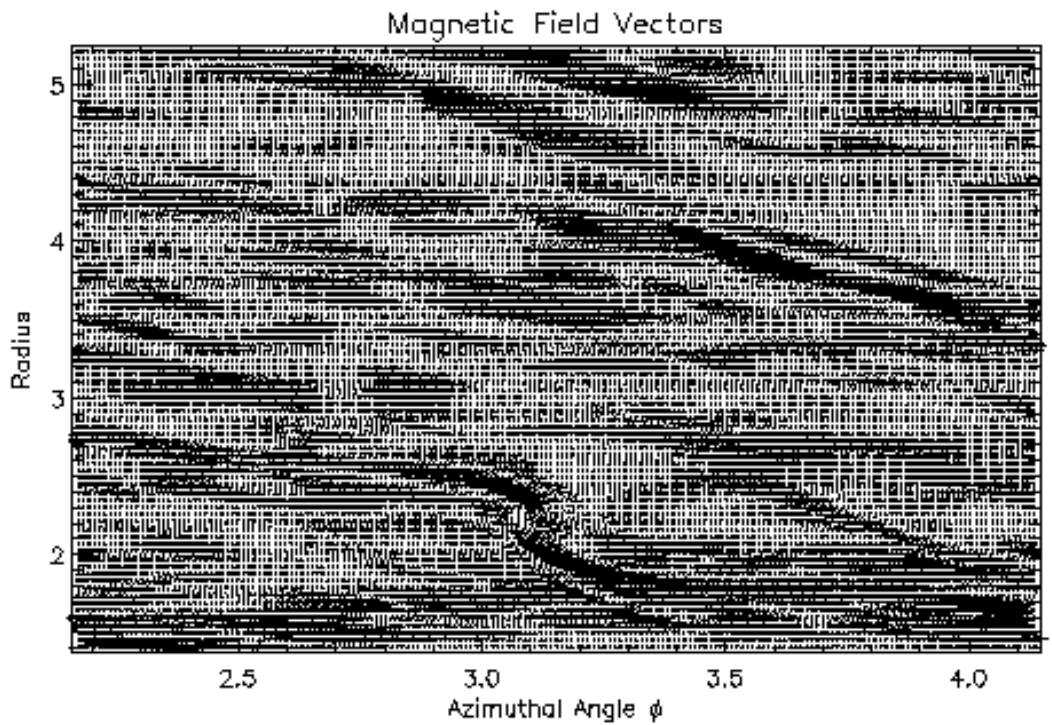,width=14cm} }
\caption[]
{This figure shows the magnetic field vectors ($B_r$, $B_{\phi}$)
in the disc midplane 
plotted in the $r$--$\phi$ plane at time $t=2040$. 
Note that the field
is compressed and ordered in the vicinity of the wakes induced by the 
protoplanet.
}
\label{fig9}
\end{figure*}

\begin{figure*}
\centerline{
\epsfig{file= 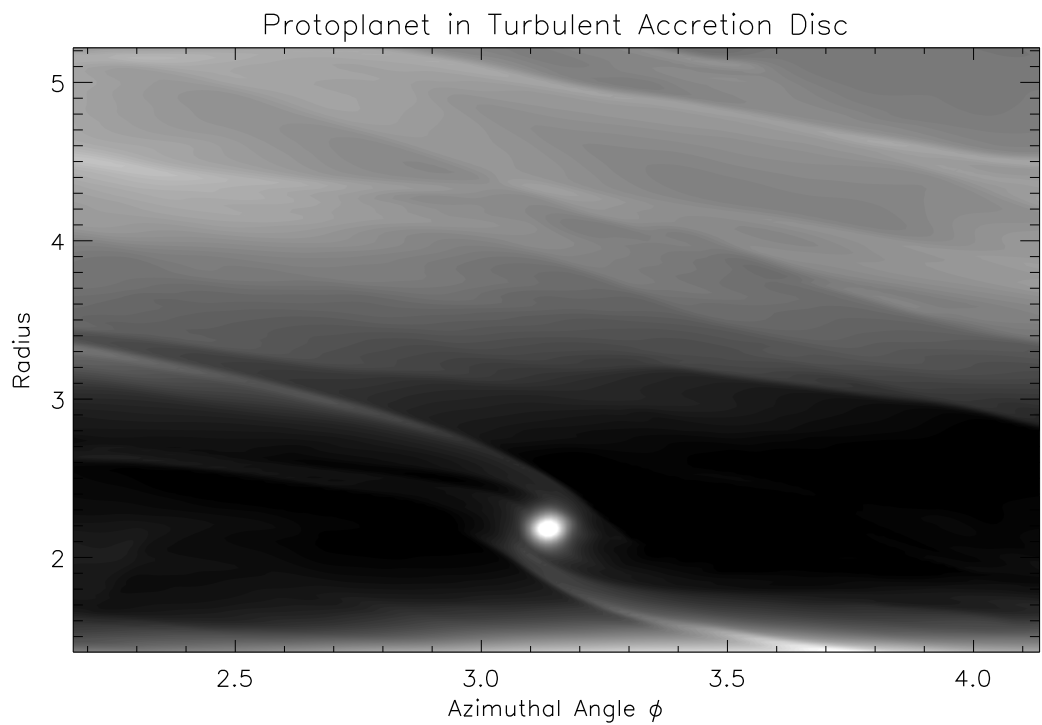,width=14cm} }
\caption[]
{This figure shows the midplane density in the vicinity of the planet
for an area that is the same as for the
magnetic field plot in figure~\ref{fig9}.
}
\label{fig10}
\end{figure*}

\begin{figure*}
\centerline{
\epsfig{file=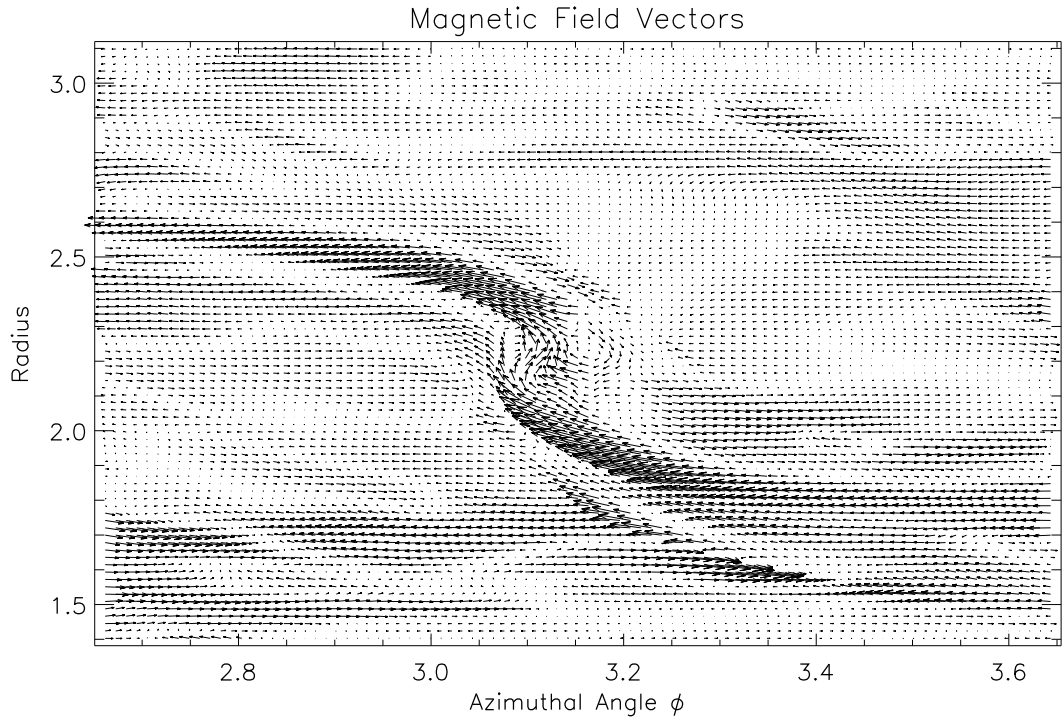,width=14cm}}
\caption[]
{This figure shows the magnetic field vectors ($B_r$, $B_{\phi}$)
in the disc midplane
plotted in the $r$--$\phi$ plane at time $t=2040$ for an area that
is somewhat smaller than that plotted in figure~\ref{fig9}.
It is clear that the field
is ordered and compressed in the vicinity of the wakes induced by the
protoplanet.
}
\label{fig11}
\end{figure*}

\begin{figure*}
\centerline{
\epsfig{file=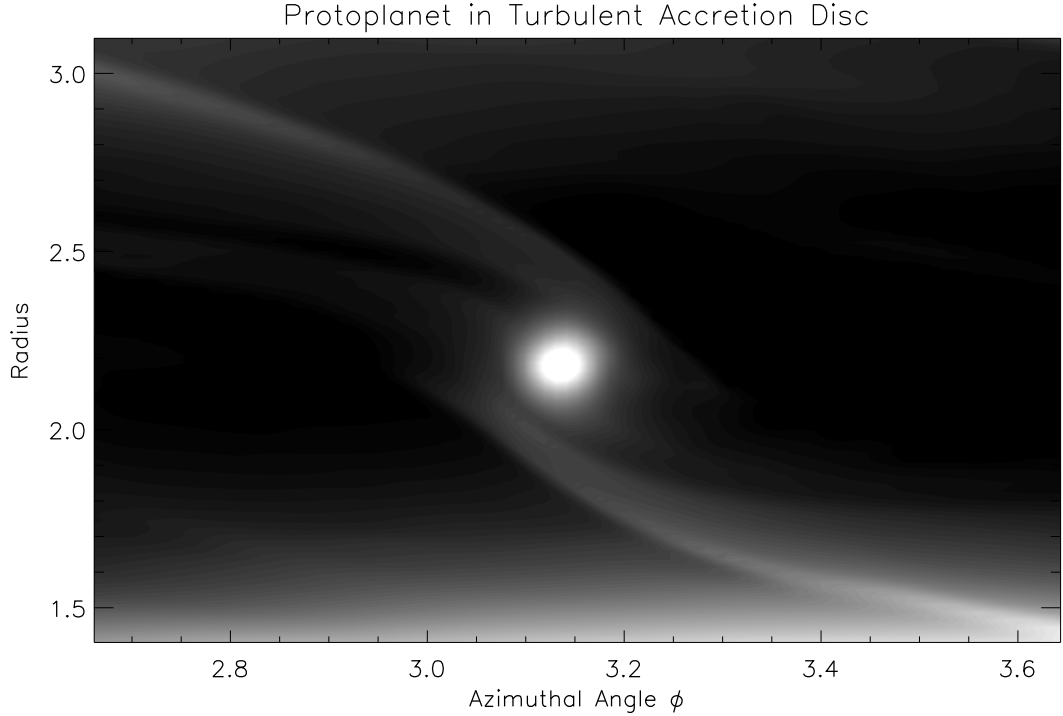,width=14cm} }
\caption[]
{This figure shows density in vicinity of planet for area equivalent to 
magnetic field plot in figure~\ref{fig11}.
}
\label{fig12}
\end{figure*}

\begin{figure*}
\centerline{
\epsfig{file=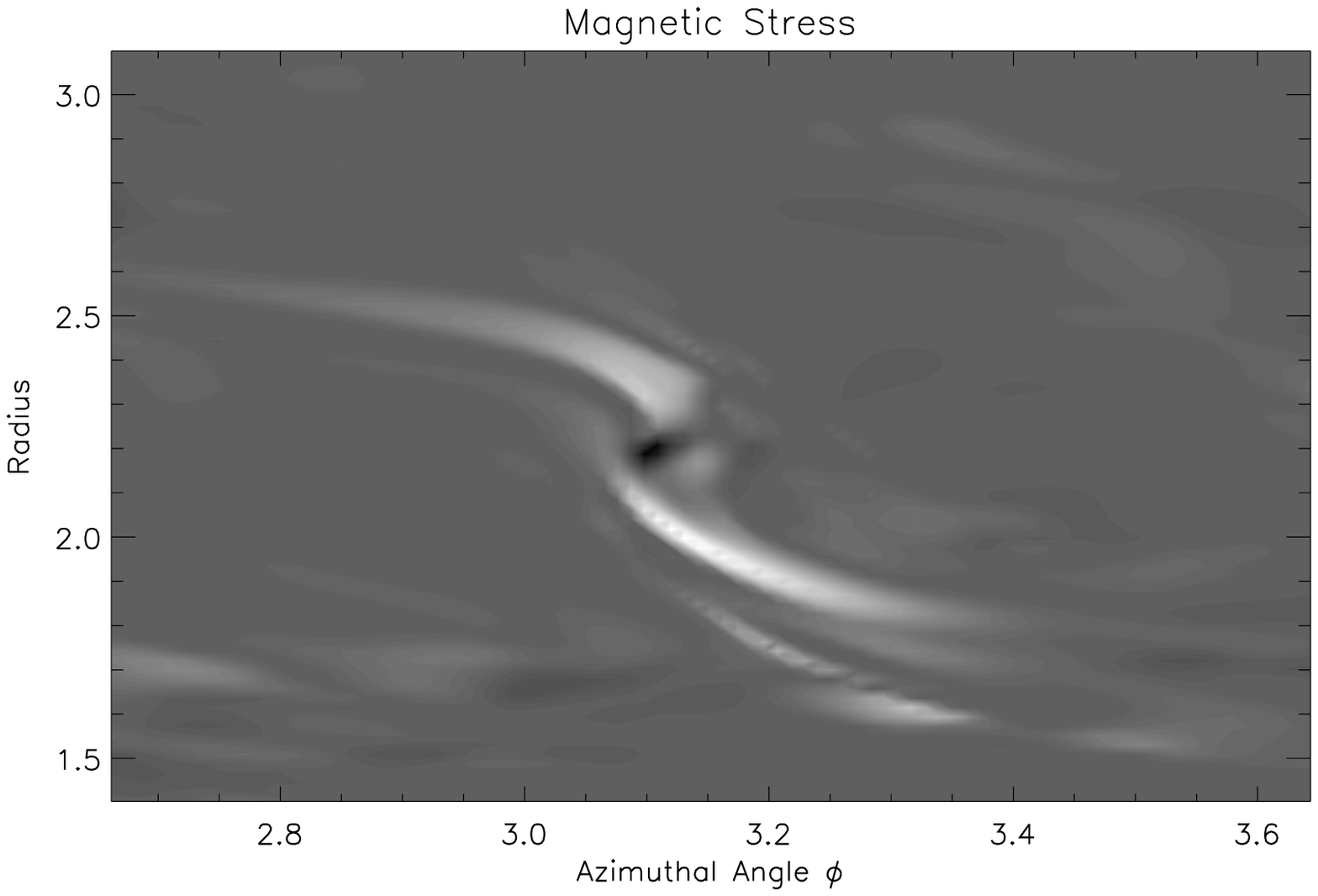,width=14cm} }
\caption[]
{This figure shows  the magnetic stress in the vicinity of the planet for 
an area equivalent to that shown in the
magnetic field plot in figure~\ref{fig11}. Note that the magnetic stress is
greatest where the magnetic field is compressed and ordered just behind the 
wakes
induced by the protoplanet.
}
\label{fig13}
\end{figure*}

In the region of the wakes the disc and its turbulence are strongly perturbed.
Magnetic field vectors in the disc midplane in the neighbourhood 
of the planet are illustrated on different size scales 
in figures~\ref{fig9} and \ref{fig11}, along with corresponding
density plots for the equivalent regions in figures~\ref{fig10} and \ref{fig12}.
An inspection of the magnetic field vectors indicates that these
tend to line up along the location of the wakes but in a somewhat
broadened region slightly behind the shocks. 
In this way an ordered structure appears to be imposed
on the flow and magnetic field by the protoplanet. 
The magnetic stress is largely communicated in these ordered regions
as shown by figure~\ref{fig13}, which shows a contour plot of the magnetic
stress in the vicinity of the planet.

\begin{figure*}
 \centerline{
 \epsfig{file=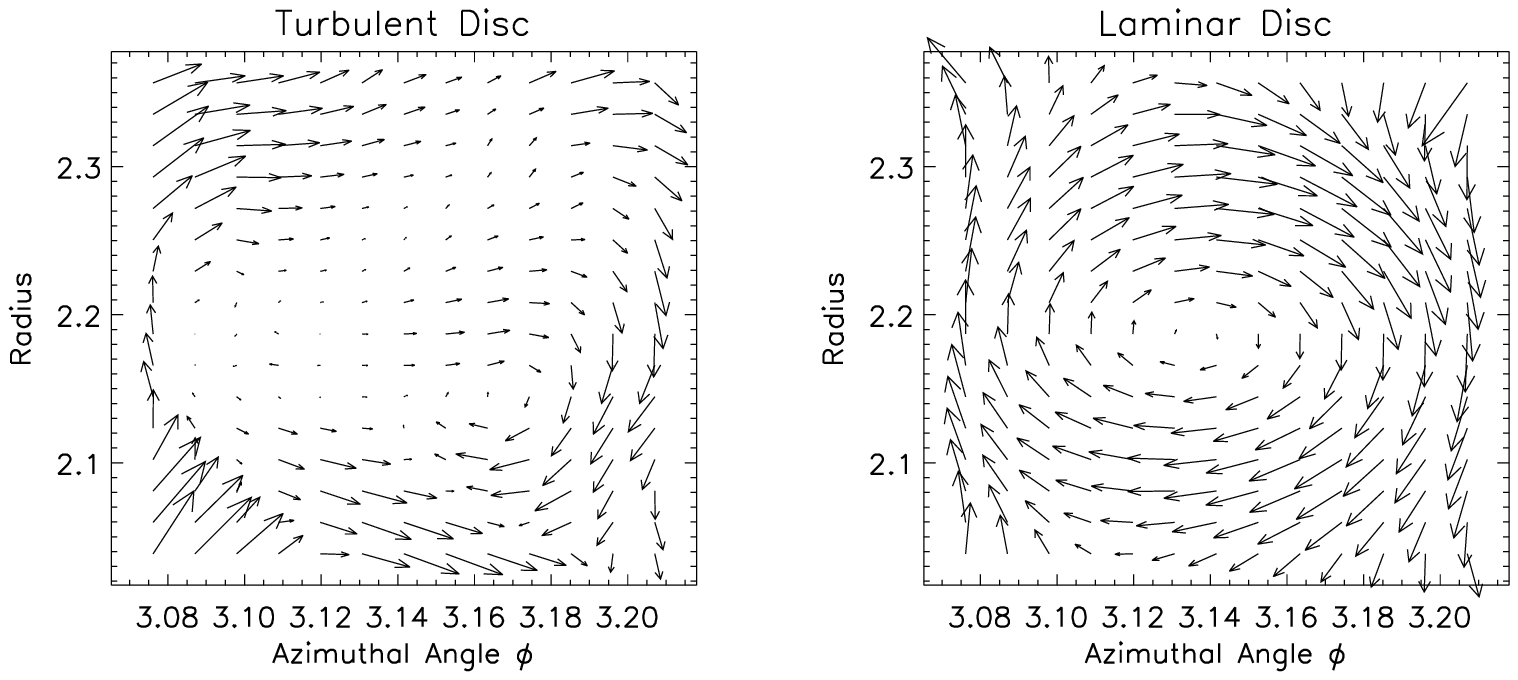,width=\textwidth}}
 \caption[]
{Velocity vectors in the disc
midplane in the Hill sphere of the planet. The left hand panel shows the
results from the magnetised disc run, the right hand panel shows a laminar disc
run with the same resolution.
Note that the usual circulating pattern that is observed
in the planet Hill sphere for non magnetic disc--planet simulations
is not apparent in the MHD turbulent case, for reasons discussed in the
text.
}
 \label{fig14}
 \end{figure*}

The velocity field within the Hill sphere of the planet is shown in the left 
hand panel of figure~\ref{fig14} for the turbulent disc. An equivalent
plot for a laminar disc is shown in the left hand panel of this figure.
The material that enters the Hill sphere of the protoplanet and becomes 
gravitationally bound to it is normally expected to circulate around it
by virtue of its angular momentum. In the calculation presented
here the protoplanet does not accrete gas, and is modelled as
a softened point mass. Consequently it is expected that material that
enters the Hill sphere will form a hydrostatic `atmosphere' around the
planet that is supported through pressure and rotation. In figure~\ref{fig14}
we would expect to see circulation occurring in a clockwise fashion for
both the MHD turbulent disc and the laminar disc.
However, it is apparent that the circulating pattern has been disrupted
in the magnetised disc, indicating
that magnetic breaking may have been responsible for this
(see section~\ref{compare}).
The expected circulation is observed in the non magnetic run performed
at the same numerical resolution as shown in the right hand panel of 
figure~\ref{fig14}.

\subsection{Comparison with laminar viscous disc models}\label{compare}

\begin{figure}
 \centerline{
 \epsfig{file=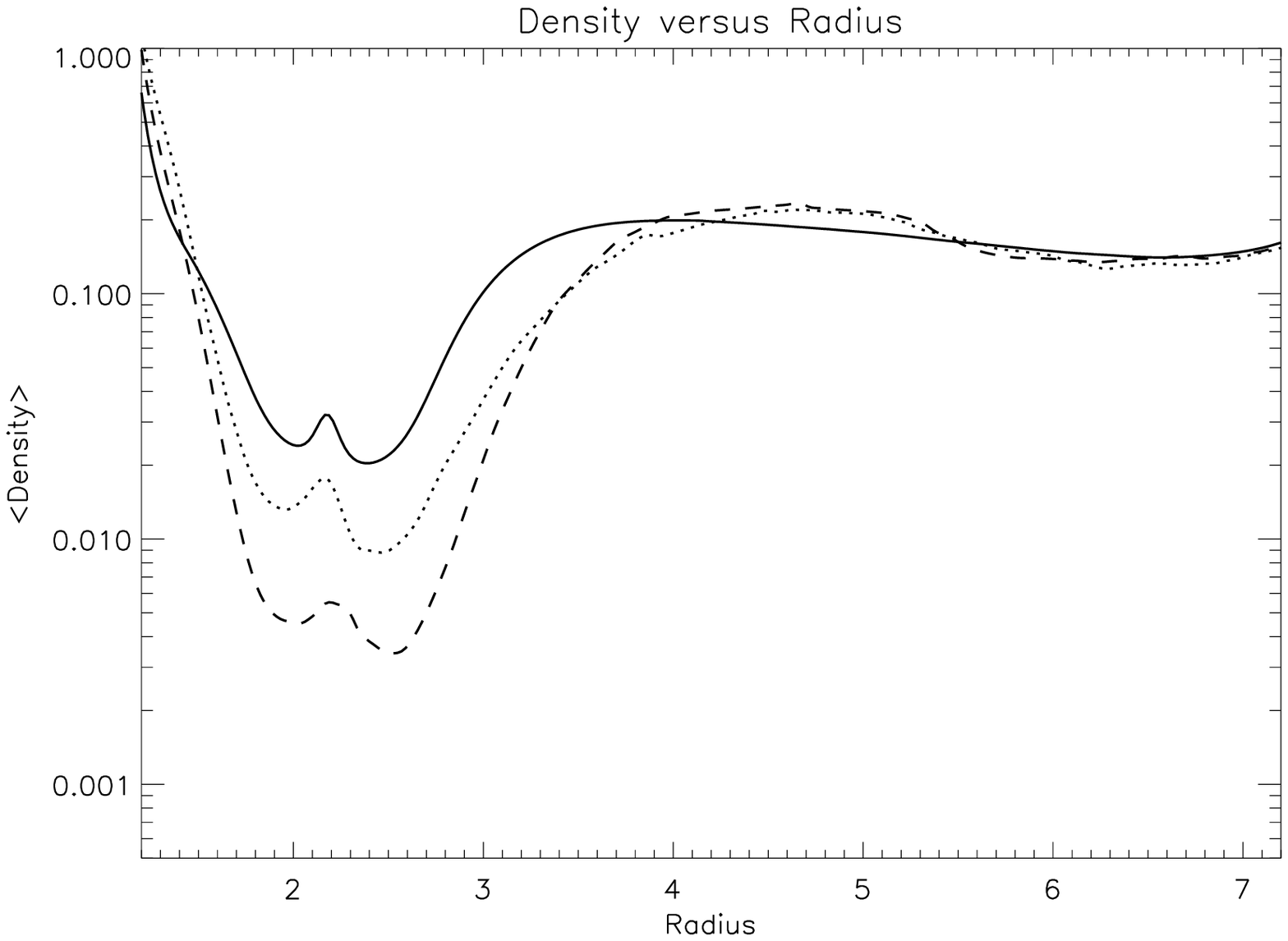,width=\columnwidth}}
 \caption[]
{This plot shows the vertical midplane density distribution for
the turbulent disc model (dotted line), a 2D laminar disc model with
the Navier-Stokes $\alpha=0$ (dashed line), and a 2D laminar disc
model with $\alpha=5 \times 10^{-3}$. It is clear that the gap for
the 2D $\alpha=0$ disc is the deepest, followed by the full turbulent
disc model, with the 2D $\alpha=5 \times 10^{-3}$ being the least deep and
wide. 
}
 \label{fig15}
 \end{figure}

\begin{figure*}
 \centerline{
 \epsfig{file=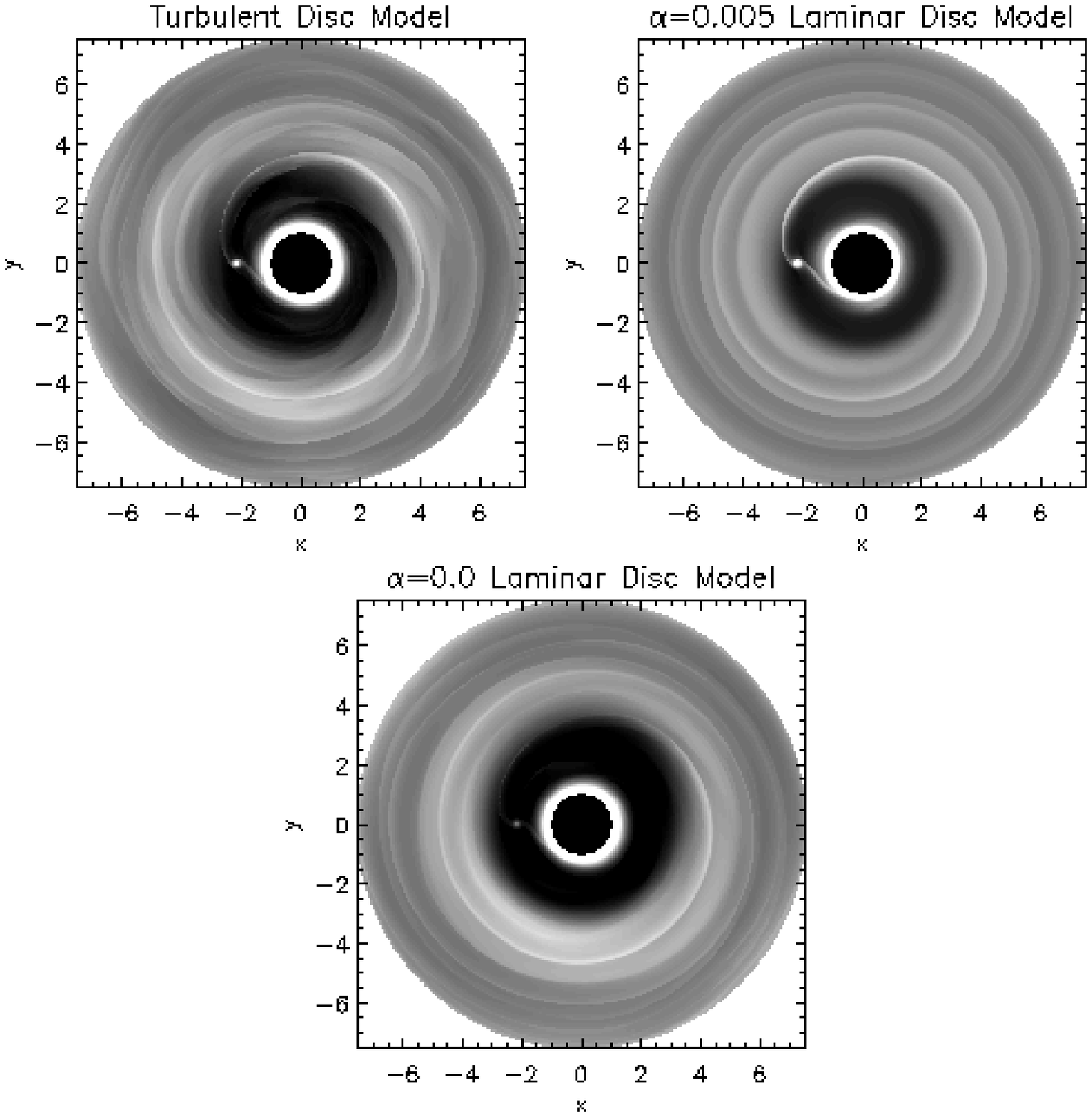,width=\textwidth}}
 \caption[]
{This plot shows the vertical midplane density distribution for
the turbulent disc model (first panel), a 2D laminar disc model with
the Navier-Stokes $\alpha=5\times 10^{-3}$ (second panel),
and a 2D laminar disc
model with $\alpha=0$ (third panel). It is apparent that the
spiral waves excited in the turbulent disc are `blurred' by the effects
of the turbulence, whereas those in the laminar disc model cases remain
more sharply defined.
}
 \label{fig15b}
 \end{figure*}

In addition to running the full MHD turbulent disc model, we have also
run some 2D laminar $\alpha$ disc models for the purposes of comparison.
A 2D model was run with $\alpha=0$, along with an $\alpha=5 \times 10^{-3}$ 
case. For initial conditions in these 2D models, we took the midplane
density distribution of the 3D turbulent model just prior to the 
addition of the planet, switched off the magnetic field, and introduced
the required value of $\alpha$ in the Navier--Stokes viscosity. The 
implementation of the Navier--Stokes viscosity is described in 
Nelson et al. (2000).  

On a qualitative level differences arise between the turbulent and
laminar disc models. The most striking visual difference is that the
spiral waves excited in the turbulent disc have a more `blurred' appearance,
whereas they appear to be more sharply defined in the laminar models
[see figures~\ref{fig8}, \ref{fig15b}, and Nelson et al. (2000)]. 
Also, an inspection of the flow topology in the close vicinity of
planet shows some differences. In the turbulent disc case, there appears
to be little sign of the usual circulating flow around the planet
for material in the Hill sphere
(see figure~\ref{fig14}),
whereas the laminar models do show the expected 
circulating flow pattern. This may
indicate that magnetic linkage between the protostellar disc
and the circumplanetary disc that forms when material flows into
the protoplanets Hill sphere leads to magnetic breaking. This effect
may be highly significant
for determining the accretion rate of gas onto the protoplanet through the 
circumplanetary disc,
and indicates that accretion rates obtained from non magnetic treatments
of the problem may be quite different. We note, however, that higher resolution
simulations will be required to examine this effect in detail, and will be
the subject of a future publication.
In addition to these qualitative differences, there are also some
important quantitative differences.

\begin{figure*}
 \centerline{
 \epsfig{file= 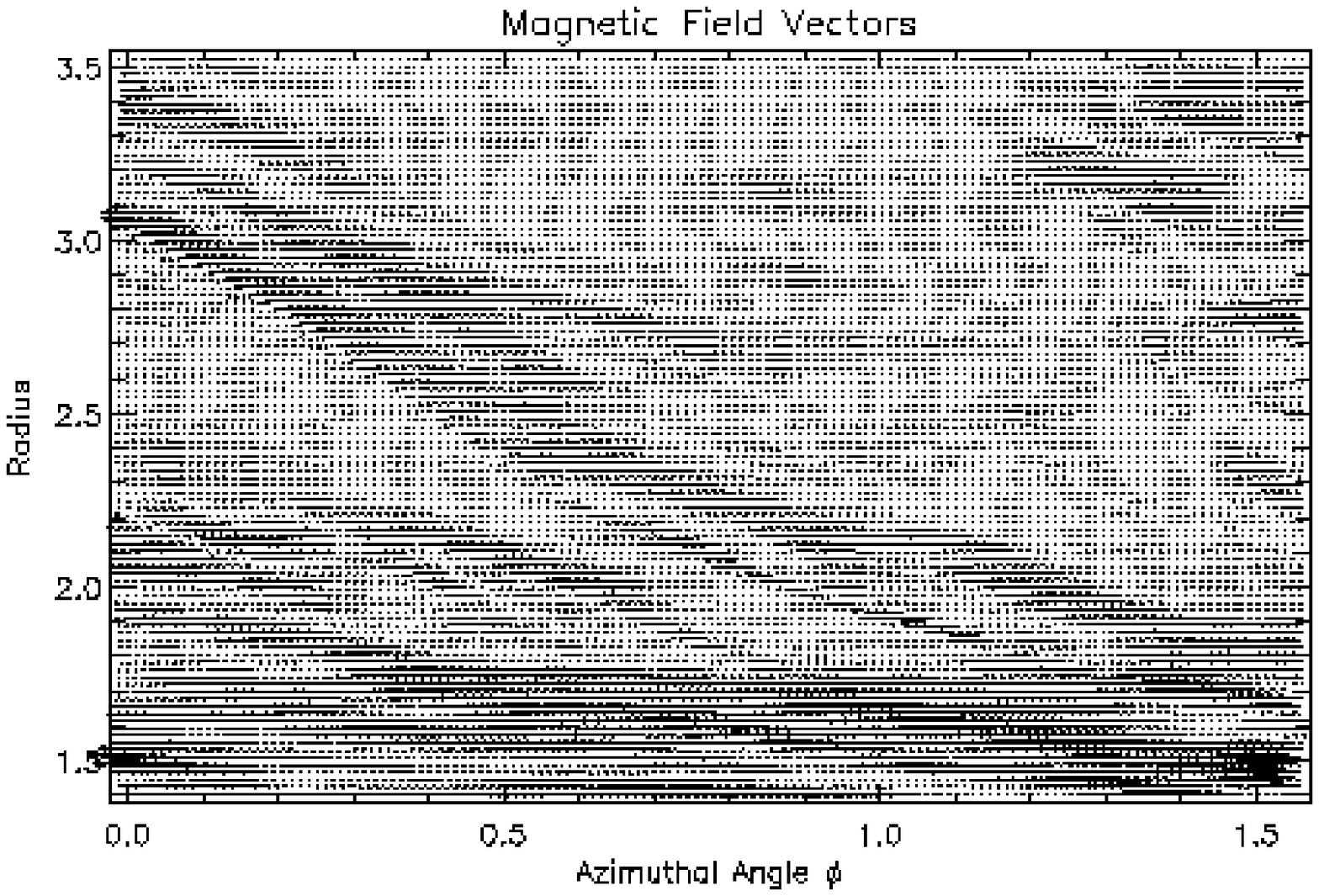,width=\textwidth}}
 \caption[]
{This plot shows magnetic field vectors ($B_r$ and $B_{\phi}$)
plotted in the $r$--$\phi$ plane of the disc vertical midplane
showing field structure in the gap region.
It is evident that magnetic field lines are able to cross the gap
(located between $r\simeq 1.8$ -- $r\simeq 3.0$) and thus transmit torques
between the two sides. This effect may be partially responsible for 
making the gap appear to  resemble one formed in a laminar disc
with smaller
effective stress parameter $\alpha$ when a protoplanet is present in
a turbulent disc.  This is because 
these magnetic stresses are able to help maintain
the gap. Note that features of this type tend to be less prevalent
in the azimuthal domain
near to the planet, but are more apparent in the regions that are 
close to $\pi$ radians
away from the planet. 
}
 \label{fig16}
 \end{figure*}

The azimuthally averaged surface density distribution
for these models, plus the midplane density distribution for the
turbulent model, is plotted in figure~\ref{fig15}. Each of the models have been
run for a total time of $\simeq 2050$ time units, corresponding to $\sim 100$
planetary orbits. It is clear that the $\alpha=0.0$ 2D model (dashed line)
has the deepest 
and widest gap, as expected. It is also apparent that the turbulent 
model (dotted line) has a deeper and wider gap than the 2D $\alpha=5 \times 10^{-3}$ model,
even though
a volume averaged estimate for the underlying turbulent disc model yields
an effective $\alpha \simeq 5 \times 10^{-3}$ (see figure 18 in paper I).
Thus, the turbulent model behaves as if it has a somewhat
smaller $\alpha$ than
reasonable estimates suggest it has. There are two probable
reasons for this. First, a Navier--Stokes viscosity with anomalous
viscosity coefficient provides a source of constantly acting friction
in the disc, such that it can induce a steady mass flow into the
gap region. The turbulence, however, does not operate as a constant
source of friction that generates steady inflow velocities. Instead
it generates large velocity fluctuations that may be much 
larger than the underlying inflow velocity arising from
the associated angular momentum transport. Results presented in paper I
indicate that a process of time averaging the turbulent velocity
field is required over long time periods before these fluctuations 
can be averaged out to reveal the underlying mass flow. The disc material
in the vicinity of the planet experiences periodic high amplitude perturbations
induced by the planet
on a time scale much shorter than the required averaging time scale, 
so that the disc response is expected to differ from that in the case of a
disc with Navier--Stokes viscosity. 
A second
 plausible reason for the apparently lower $\alpha$ is
that the existence of the magnetic field in the turbulent disc
allows for field lines to connect across the gap region, and to enable
angular momentum transport across the gap. In this way the magnetic
field actually helps the planet to maintain the gap. Figure~\ref{fig16}
shows an example of how magnetic structures are able to extend across
the gap and thus maintain communication across the gap region. The stresses
associated with these structures lead to outwards angular momentum transport,
and thus assist in maintaining the gap.
However, it should be borne in mind that the above
remarks apply to one particular simulation involving a rather thick disc
with $H/r = 0.1$ and a 5 Jupiter mass  protoplanet.
The ability of ordered magnetic field to traverse the gap 
may be much less effective 
in a thinner disc in which the gap structure tends to be
sharper and deeper. Also the existence of significant disc material
interior to the planet's orbit may assist this process, and in a model
with an extensive  inner cavity it may be less efficient.
Nonetheless the differences in gap structure found here
suggest that the accretion rate
onto a protoplanet in an MHD turbulent disc is likely to be less
than previous estimates based on laminar  $\alpha$ disc models indicate
[e.g. Bryden et al. (1999); Kley (1999); Lubow, Seibert,
\& Artymowicz (1999); Nelson et al. (2000); D'Angelo, Henning, \& Kley (2002)].

\section{Discussion and conclusions} \label{conclusions}
We have presented a global
MHD simulation of a turbulent  disc interacting with  a giant
protoplanet in a fixed circular orbit.
The disc model that we considered (described in detail in paper I)
had $H/r=0.1$, and in the absence of the protoplanet
a volume  and time
averaged value of the stress parameter $\alpha = 5 \times 10^{-3}$.
We considered a protoplanet
with 5 Jupiter masses which from previous work was 
expected to open and maintain a significant gap.
Although 
the available
computational resources have limited the parameters
that could be adopted,
we expect the
essence of the disc--protoplanet interaction to be  captured.

 Using appropriate time averaged stress parameters and radial
velocities, many of the phenomena seen in two dimensional
laminar disc simulations could be demonstrated.

Spiral waves  that produced an outward angular momentum
flux through Reynolds' stresses were launched by the protoplanet.
As expected the existence of the  waves reduced the magnitude of  
the stresses provided by the MHD turbulence so  reducing
the associated  outward angular momentum flow  and 
 thus  maintaining  the gap.
A net outward angular momentum flow from the planet orbit
to the disc was seen which leads to inward migration
of the protoplanet orbit on a time scale expected from type II migration theory.

When compared with laminar disc models,
with the same estimated $\alpha,$
the gap in the turbulent disc was found to be deeper indicating that 
the turbulent
disc behaved as if it possessed a smaller $\alpha.$
This was in spite of the fact that the turbulent disc
appeared more effective at disrupting and dissipating the spiral waves.
This behaviour may occur for two basic reasons. The first is that
the turbulence does not provide a source of constantly acting friction
in the disc, unlike a Navier--Stokes viscosity, and time averaging for
significant periods is required before velocity fluctuations cancel
out to reveal the underlying mass flow. The disc is periodically perturbed by
the planet on time scales shorter than those required for the time averaging,
indicating that the disc response is likely to differ from a disc with
Navier--Stokes viscosity operating.
The second reason is that 
ordered magnetic structures where found to be able cross the gap at azimuthal
locations that were displaced from the planet by $\approx \pi$.
Such a connection would enable an outward flux of angular momentum to flow
across the gap and help to maintain it independently
of the planetary torques, so, as far as gap
properties are concerned making the disc appear
to behave as if  there was a smaller $\alpha$.
We also found no evidence for a persistent mass flow
through the gap from the outer to inner disc but we emphasize that
this may change if an extensive deep inner disc cavity can
be produced or if the disc is thinner.
On the other hand it may persist for smaller protoplanets
which make weaker gaps and inner cavities.

The behaviour of the material bound to the planet was found to
behave differently in the case of a magnetised disc, with the circulating
material
within the planetary Hill sphere apparently experiencing magnetic breaking.
This has implications for the structure and evolution of the circumplanetary
disc that is expected to form during the latter stages of gas giant planet 
formation, and the mass accretion rate onto giant protoplanets.
These issues require detailed investigation
through additional simulations, and will be the subject of future publications.

\subsection{Acknowledgements} 

We  acknowledge  John Hawley, Steve Balbus, Adriane Steinacker, and
Caroline Terquem for useful
discussions.
The computations reported here were performed using the UK Astrophysical 
Fluids Facility (UKAFF).

{}
\end{document}